\documentstyle[12pt]{article}
\begin{document}
SPACE-TIME TORSION AND THE ROTATION OF GALAXIES\\
\bigskip

M.L. Fil'chenkov\\
Alexander Friedmann Laboratory for Theoretical Physics,\\
26-9 Konstantinov Street, Moscow 129278, Russia\\
E-mail: fil@agmar.ru\\

\begin{abstract}
Torsion effects, including a spin precession in the torsion field, are
considered. Some properties of neutrinos in cosmology are discussed. In the
framework of Trautman's cosmological model with torsion estimated is a
specific angular momentum of initial perturbations which proved to be of the
order of the observable specific rotational moment for spiral galaxies. The
results obtained are compared with those from the theories of potential and
vortical perturbations in which rotation of galaxies is predicted.
\end{abstract}

\section{Introduction}
While considering classical effects and quantum systems in strong
gravitational fields, theories alternative to General Relativity can hardly be
passed over in silence. On the other hand, the modern structures (stars,
galaxies, clusters of galaxies) can only be understood from cosmological
outlook, viz. considering the quantum early Universe.

Below we present a model with spin and torsion which can explain the origin
of galactic rotation on the basis of quantum torsion effects. The early
Universe is a quantum system whose hierarchical structure is related to
initial perturbations. The latter, in turn, are formed from quantum
fluctuations whose specific angular momenta are $LS$-coupled on a small scale
and $jj$-coupled on a large one. The $LS$-coupling is classically interpreted
as a precession of $L$ and $S$ about $J$. This is a mechanism of the angular
momentum transfer due to the spin precession in the torsion field. Thus as a
source of galactic rotation may be massive particles with spin comprising the
so-called dark matter. The predicted specific angular momenta proves to be
of the order of the observable ones for spiral galaxies. The results obtained
are compared with those from the theories of potential and vortical
perturbations in which rotation of galaxies is predicted.

\section{Torsion Effects}
Immediately after General Relativity had been created, there appreared its
generalizations, viz. Einstein-Cartan's (Cartan, 1922), Kaluza-Klein's
(Kaluza, 1921) and  Weyl's (Weyl, 1918) theories. In particular, in
Einstein-Cartan's theory (ECT) (Trautman, 1979; Ivanenko et al., 1985a;
Ivanenko and Sardanashvili, 1985b; Ponomarev et al., 1985; Rodichev, 1974;
Sabata, 1994)  considered is a nonsymmetric connection (torsion)
$$
Q_{\mu\nu}{}^{\lambda}=\frac{1}{2}(\Gamma_{\mu\nu}{}^{\lambda}-\Gamma_{\nu\mu}
{}^{\lambda}).                                              \eqno(1)
$$
The torsion contributes to the energy-momentum tensor of the spinning matter
being its source, which results in eliminating the singularities of a
gravitational field, e.g. in cosmology. Really, the energy-momentum tensor of
a spin liquid has the form (Ivanenko et al., 1985a)
$$
T_{\mu\nu}{}^{eff}=u_{\mu}u_{\nu}(p+\rho-2s^2)-g_{\mu\nu}(p-s^2), \eqno(2)
$$
with $s^2=s_{\mu\nu}s^{\mu\nu}$,
where the spin $s$ leads to an effective negative pressure and eliminates the
singularity.

On the other hand, writing down the Lagrangian of the spinor field in
Riemann-Cartan's space, one can obtain a nonlinear spinor equation coinciding
with Ivanenko-Heisenberg's (Ivanenko and Sardanashvili, 1985b)
$$
\gamma^{\mu}D^{\mu}\psi+\frac{3}{8\varepsilon}(\psi\gamma_{\mu}\gamma_{5}\psi)
\gamma^{\mu}\gamma^{5}\psi=0                             \eqno(3)
$$
where $\varepsilon$ is a constant of interaction with the torsion field.\\
A relation of the torsion field to a nonlinearity has first been revealed by
V.I. Rodichev (1961) for the case of the absence of a gravitational field.

Finally, the torsion $Q$ leads to a spin precession (Yefremov, 1980)
$$
\frac{d{\bf s}}{dt}=c[{\bf Q}{\bf s}]                            \eqno(4)
$$
where ${\bf s}$ is the spin vector $(|{\bf s}|=s)$, ${\bf Q}$ is the
"polarized" torsion vector.
$$
Q=c\kappa S,                                                       \eqno(5)
$$
where $\kappa=\frac{8\pi G}{c^4}$, $G$ is the gravitational constant, $c$ is
the velocity of light, $S$ is the matter spin density creating the torsion
$Q$.

It should be noted that Einstein-Cartan's theory is now a universally
recognized generalization of General Relativity to the case of taking account
of the spin of matter in the early Universe (Ivanenko et al., 1985a).

\section{Comments on Massive Neutrinos in Cosmology}

The modern upper limits to the neutrino mass (Boehm and Vogel, 1987) do not
contradict a closed model of the Universe wherein it is neutrinos that
determine the space-time structure on a cosmological scale, because the
neutrino background by 1-2 orders exceeds the average densitity of the matter
being observed in galaxies (Dolgov et al., 1988). Since the spinor fields
describing the neutrinos are a natural source of torsion, it is not
unreasonable to consider cosmology in the framework of ECT.
For the massless neutrino there exists a difficulty consisting in Weyl's
equation having only a trivial solution for spherically symmetric
configurations both in the framework of GR (Audretsch, 1972) and ECT
(Kuchowicz, 1974).
Another paradox of the massless neutrino is an appearance of "ghosts", i.e.
such solutions for which the energy-momentum tensor is identically equal to
zero, whereas the field and the current are nonzero (Edmonds, 1976). The
paradoxes of the massless neutrino suggest an idea of these difficultieas
being related to the assumption that the neutrino is massless. Whence it
follows that the consideration of the massive neutrino is, at least, not
unreasonable.

\section{Cosmological Model with Torsion}

The cosmology with torsion has been investigated by A. Trautman (1973)
who arrived at the conclusion that the torsion eliminates the singularity and
stops the collapse (in the case of a closed model) at the minimum radius\\
$R\sim 1$ cm, with the matter density $\rho\sim 10^{55}$ gcm${}^{-3}$. These
values are obtained assuming that the source of torsion is $10^{80}$ nucleons
with polarized spins. In the framework of ECT the following formulae were
used for Friedmann's universe:
$$
R_{min}=\left(\frac{3G\hbar^2N}{8mc^4}\right)^{\frac{1}{3}},      \eqno(6)
$$
$$
\rho_{max}=\frac{4m^2c^4}{3\pi^2G\hbar^2}                          \eqno(7)
$$
where $N$ is the number of nucleons, $m$ is their rest mass. Formulae (6)-(7)
also remain valid for a chaotic spin distribution with $<{\bf S}>=0$ and
$<S^2>\ne 0$ (Ponomarev et al., 1985).

If we assume the neutrinos with the rest energy $m_{\nu}c^2\sim$ 35 eV
(Lyubimov, 1980) to be a source of torsion, then we shall obtain the values of
parameters as follows:\\
$R_{min}\simeq 2\cdot 10^5$ cm, $\rho_{max}\simeq 4\cdot 10^{39}$ gcm${}^{-3}
$. It is easy to see that the separation of neutrinos $l_{min}=(\rho_{max}/m_
{\nu})^{-1/3}\sim 10^{-24}$ cm which much less than the Compton wavelength
$\lambda=\frac{\hbar}{m_{\nu}c}\sim 10^{-6}$ cm. This means that the problem
should be considered at least in terms of quantum theory.

\section{Estimation of the Specific Angular Momentum}

From quantum mechanics it is known that for light atoms there occurs the
Russell-Saunders or $LS$-coupling (Landau and Lifshitz, 1963) when\\
$$
L=\sum_{i} ^{} l_{i},\quad S=\sum_{i} ^{} s_{i},\quad J=L+S.
$$
This is called a vector model of the atom to which in terms of classical
mechanics corresponds a precession of the vectors $L$ and $S$ about the total
angular momentum vector $J$ (Blokhintsev, 1976).

On the other hand, from ECT it is known that the spin of a test body precesses
in the torsion field (see Sec. 1) with the frequency
$$
\Omega_{pr}=cQ.                                                   \eqno(8)
$$
From (5) for $S=\frac{\hbar}{2}n$ we obtain $Q=\frac{\hbar}{2}c\kappa n$,
where $n$ is the density of the number of particles with the spin $\frac
{\hbar}{2}$. For the precession frequency we obtain
$$
\Omega_{pr}=\frac{4\pi\hbar Gn}{c^2}.                              \eqno(9)
$$
Notice that the quantity of the spin of a test body does not enter in formula
(9).

From classical mechanics it is known (Landau and Lifshitz, 1978) that the
precession of a symmetric top occurs about the direction of its total angular
momentum with the frequency
$$
\Omega_{pr}=\frac{J}{I}                                            \eqno(10)
$$
where $I$ is the moment of inertia.
Using the relation
$$
I\simeq MR^2,                                                       \eqno(11)
$$
we obtain for the specific angular momentum of the top
$$
\frac{J}{M}\simeq\Omega_{pr}r^2                                    \eqno(12)
$$
where $M$ is the top mass, $R$ is its effective radius.

Hence, if we assume that the initial perturbations corresponding to
protogalaxies with the spin moment $S$ and the orbital moment $L$ could be
added (in terms of quantum mechanics), i.e. could precess (in terms of
classical mechanics) about the total angular momentum $J$ as well as the
uncompensated spin ${\bf s}$ of a test body precesses about the "polarized"
torsion ${\bf Q}$ being created by the particles having a spin, then to
estimate their specific angular momentum, we shall be able to use formula
(12), where the torsion $Q$ plays the role of the total angular momentum
$J$, i.e. in ECT by analogy with the $LS$-coupling we have $Q=\frac{L+S}{cI}$.
This means that the spins of small perturbations are added into the total
torsion vectors of protogalaxies.  The specific angular momentum in formula
(12) is related to the total torsion vector that in a selfconsistent system
comprises uncompensated spin angular momenta of the initial perturbations
being their source at the same time. Thus from (8) and (10) we have
$$
J=cIQ                                                           \eqno(13)
$$
where
$$
I=\rho\int_0^a r^2\,dV,                                          \eqno(14)
$$
$V=2\pi^2r^3$ for a closed model.\\
Hence
$$
I=\frac{3}{5}a^2M                                                \eqno(15)
$$
where $M=2\pi^2\rho a^3$.\\
For the specific angular momentum we obtain the formula
$$
\frac{J}{M}=\frac{12\pi\hbar G}{5c^2}na^2=\frac{3}{5}\Omega_{pr}a^2 \eqno(16)
$$
similar to (12).

The neutrino background, on the one hand, is a source of the gravitational
field,  and on the other hand, is a quantum system, at least for the early
Universe. Hence a correct description of its behaviour is possible only in
the framework of quantum theory. To estimate the specific angular momentum,
we shall consider the spin precession of initial perturbations to occur in
the neighbourhood of the minimum radius of the Universe. In formula (16) we
assume that $n=n_{max}$,\quad $a=R_{min}$. From (6), (7) we have
$$
n_{max}=\frac{4mc^4}{3\pi^2 G\hbar^2},\quad R_{min}=\left(\frac{3G\hbar^2N}
{8mc^4}\right)^{\frac{1}{3}}
$$
where $n_{max}=\frac{\rho_{max}}{m}$.\\
Hence
$$
\frac{J}{M}=\frac{16mc^2}{5\pi\hbar}\left(\frac{3G\hbar^2N}{8mc^4}\right)^
{\frac{2}{3}}.                                                 \eqno(17)
$$
Using the formulae (Zel'dovich and Novikov, 1975)
$$
N=nV,\quad V=2\pi^2a_0{}^3,\quad a_0=\frac{c}{H_0}\frac{1}{\sqrt{\Omega-1}},
\quad \Omega=\frac{8\pi Gmn}{3H_0{}^2},
$$
where $a_0$ is the scale factor, $H_0$ is the Hubble constant and $\Omega$ is
the average density in units of the critical one (the index "0" corresponds
to the present epoch), we can
obtain
$$
N=\frac{3\pi c^3}{4H_0Gm}\frac{\Omega}{(\Omega-1)^3}           \eqno(18)
$$
and finally express $\frac{J}{M}$ in terms of $\Omega$ as follows:
$$
\frac{J}{M}=\frac{6}{5}\sqrt[3]{\frac{12}{\pi}}\frac{1}{\Omega-1}\left(\frac
{c\Omega}{H_0\lambda}\right)^{\frac{2}{3}}\frac{S}{M}            \eqno(19)
$$
where the Compton wavelength of spin particles $\lambda=\frac{\hbar}{mc}$,
the specific spin moment $\frac{S}{M}=\frac{\hbar}{2m}$.

From this formula it follows that $J\gg S$ for $\Omega-1\ll 1$ since $\frac{c}
{H_0\lambda}\gg 1$. Note that $\frac{J}{S}$ does not depend on $G$ which
enters only via $\Omega$ being of the order of unity. Hence, if $J=L+S$,
then $J\approx L$. This means an angular momentum transfer due to the spin
precession in the torsion field. For $\Omega-1\ll 1$ the required $\frac{J}
{M}$ is always achievable by tuning the mass of the particle being a source
of the spin. For example, for $\Omega-1=10^{-2}$\quad $m_{\nu}c^2=13$ eV, $n_{
\nu}=450$ cm${}^{-3}$ (a massive neutrino), $H_0=75$ km$\cdot s{}^{-1}$Mps${}^
{-1}$ we obtain $\frac{J}{M}=2\cdot 10^{29}$ cm${}^2$s${}^{-1}$ which is close
to the corresponding value for spiral galaxies (for our Galaxy $K=5\cdot 10^
{29}$cm${}^2$s${}^{-1}$ (Ozernoy, 1978) since the specific angular momenta of
protogalaxies are conserved and equal to those of galaxies observable at
present. Angular momenta of protogalaxies are $jj$-coupled into the total
momentum of a closed Universe equal to zero.

The observed anisotropy of the microwave background radiation sets an upper
bound on the density and velocity perturbations of clusters but not galaxies.
For the clusters of galaxies the averaged angular momenta are close to zero
and do not contribute to the observed $\frac{\Delta T}{T}\sim 10^{-5}$, i.e.
only due to density perturbations.

\section{Conclusion}

We have shown that the problem of the origin of galactic rotation is solvable
in the framework of a cosmology taking account of spin and torsion. It has
partly been solved in the theories of potential and vortical perturbations
(Ozernoy, 1978; Gurevivch and Chernin, 1978; Vorontsov-Velyaminov, 1972). In
the latter case considered are chaotic supersonic turbulent motions of the
matter density and velocity perturbations having spin and orbital rotations
about each other. We see that this picture qualitatively resembles that
considered above in the framework of ECT. In this connexion we notice paper
(Soares, 1981) where  generation of a macroscopic asymmetry of neutrinos due
to the macroscopic vortical field of matter is considered, i.e this is a
process in some sense inverse to ours.

\section{Acknowledgement}

I am grateful to A.D. Chernin for helpful discussions.\\

{\bf References}\\
Audretsch J. (1972) {\em Lett. Nuovo Cimento}, {\bf 4}, No. 9, 339.\\
Blokhintsev D.I. (1976) {\em Foundations of Quantum Mechanics}, Nauka, Moscow.
\\
Boehm F. and Vogel P. (1987) {\em Physics of Massive Neutrinos}, Cambridge
University Press, Cambridge(UK).\\
Cartan M.E. (1922) {\em Compt. Rend.}, {\bf 174}, 593.\\
Dolgov A.D., Zel'dovich Ya.B. and Sazhin M.V. (1988) {\em Cosmology of the
Early Universe}, Moscow University Publishers, Moscow.\\
Edmonds T.D. (1976) {\em Lett. Nuovo Cimento}, {\bf 17}, No. 1, 34.\\
Gurevich L.E. and Chernin A.O. (1978) {\em Introduction to Cosmogony}, Nauka,
Moscow.\\
Ivanenko D.D., Pronin P.I. and Sardanashvili G.A. (1985a) {\em Gauge
Gravitation Theory}, Moscow University Publishers, Moscow.\\
Ivanenko D.D. and Sardanashvili G.A. (1985b) {\em Gravitation}, Naukova Dumka,
Kiev.\\
Kaluza T. (1921) {\em Sitzungsber. d. Berl. Akad.}, 966.\\
Kuchowicz B. (1974) {\em Phys. Lett. A}, {\bf 50}, No. 4, 267.\\
Landau L.D. and Lifshitz E.M. (1963) {\em Quantum Mechanics. Nonrelativistic
Theory}, Nauka, Moscow.\\
Landau L.D. and Lifshitz E.M. (1978) {\em Mechanics}, Nauka, Moscow.\\
Lyubimov V.A. (1980) {\em Atom. Energ.}, {\bf 49}, No. 3, 349.\\
Ozernoy L.M. ( 1978) In: {\em "Origin and Evolution of Galaxies and Stars"},
Pikel'ner S.B. (ed.), Nauka, Moscow, p. 105.\\
Ponomarev V.N., Barvinsky A.O. and Obukhov Yu.N., (1985) {\em Geometrodynamic
Methods in a Gauge Approach to the Gravitation Interaction Theory},
Energoatomizdat, Moscow.\\
Rodichev V.I., (1974) {\em Gravitation Theory in an Orthogonal Frame}, Nauka,
Moscow.\\
Rodichev V.I. (1961) {\em Zhurn. Eksper. i Teoret. Fiz.}, {\bf 40}, 1469.\\
Sabata de V., (1994) {\em G. Astr. (Roma)}, {\bf 20}, No. 2, 23.\\
Soares I.D. (1981) {\em Phys. Rev. D}, {\bf 23}, No. 2, 272.\\
Trautman A. (1979), {\em Symposia Mathematica}, {\bf 12}, No. 1, 139.\\
Trautman A.  (1973) {\em Nature. Phys. Sci.}, {\bf 242}, No. 114, 7.\\
Vorontsov-Velyaminov B.A. (1972) {\em Extragalactic Astronomy}, Nauka, Moscow.\\
Weyl H. (1918) {\em Sitzungsber. d. Berl. Akad.}, 465\\
Yefremov A.P. (1980) {\em Izv, VUZov. Fizika}, No. 8, 84.\\
Zel'dovich Ya.B. and Novikov I.D. (1975) {\em Structure and Evolution of the
Universe}, Nauka, Moscow.\\

\end{document}